\documentclass[11pt]{article}
\usepackage{times}
\usepackage{sidecap_mod,xspace}
\usepackage{geometry}
\usepackage{floatrow}
\usepackage{float}
\usepackage{hyperref}
\usepackage{graphics,graphicx}
\usepackage{subcaption}
\usepackage{epsf,wrapfig}

\usepackage{natbib,bm}

\usepackage{paralist}
\let\olditem\item
\renewenvironment{thebibliography}[1]{%
  \section*{\refname}
  \let\par\relax
  \renewcommand{\item}[1][]{\olditem[\textbullet]}%
  \inparaenum}{\endinparaenum}

\newcounter{aff}

\DeclareGraphicsExtensions{.jpg,.pdf,.png,.eps}

\usepackage{caption}
\usepackage{xcolor}

\usepackage{tcolorbox}

\bibpunct{(}{)}{;}{a}{}{,}

\usepackage{soul}

\definecolor{darkcyan}{rgb}{0.0, 0.0, 0.7}

\definecolor{boxcolor}{RGB}{193, 229, 255}
\definecolor{figboxcolor}{RGB}{193, 229, 255}
\definecolor{boxcolorgray}{RGB}{227, 227, 227}
\definecolor{darkcyan}{rgb}{0.0, 0.0, 0.7}

\sethlcolor{boxcolorgray}

\def\gtrsim{\lower 2pt \hbox{$\, \buildrel {\scriptstyle >}\over
{\scriptstyle \sim}\,$}}
\def\lesssim{\lower 2pt \hbox{$\, \buildrel {\scriptstyle <}\over
{\scriptstyle \sim}\,$}}

\def\approxlt{\lower.2em\hbox{$\buildrel < \over \sim$}}
\def\approxgt{\lower.2em\hbox{$\buildrel > \over \sim$}}

\newcommand{\um}{\ensuremath{\mu\rm{m}}\xspace}

\newcommand{\Msun}{\ensuremath{\rm{M}_\odot}\xspace}

\newcommand{\kms}{\ensuremath{\rm{km\,s}^{-1}}\xspace}

\newcommand{\sigsfr}{\ensuremath{\Sigma_{\rm{SFR}}}\xspace}

\newcommand{\Mstar}{\ensuremath{M_{\rm{star}}}\xspace}

\newcommand{\oiii}{[O{\scriptsize III}]\xspace}
\newcommand{\cii}{[C{\scriptsize II}]\xspace}
\newcommand{\ci}{[C{\scriptsize I}]\xspace}

\def\msun{\rm M_{\odot}}

\def\be{\begin{equation}}
\def\ee{\end{equation}}
\def\bea{\begin{eqnarray*}}
\def\eea{\end{eqnarray*}}

\def\oiii{[O{\sc iii}]}

\def\jref@jnl#1{{\rm#1}}
\def\aj{\jref@jnl{AJ}}			
\def\araa{\jref@jnl{ARA\&A}}		
\def\apj{\jref@jnl{ApJ}}			
\def\apjl{\jref@jnl{ApJ}}		
\def\apjs{\jref@jnl{ApJS}}		
\def\ao{\jref@jnl{Appl.~Opt.}}		
\def\apss{\jref@jnl{Ap\&SS}}		
\def\aap{\jref@jnl{A\&A}}		
\def\aapr{\jref@jnl{A\&A~Rev.}}		
\def\aaps{\jref@jnl{A\&AS}}		
\def\azh{\jref@jnl{AZh}}			
\def\baas{\jref@jnl{BAAS}}		
\def\jrasc{\jref@jnl{JRASC}}		
\def\memras{\jref@jnl{MmRAS}}		
\def\mnras{\jref@jnl{MNRAS}}		
\def\pra{\jref@jnl{Phys.~Rev.~A}}	
\def\prb{\jref@jnl{Phys.~Rev.~B}}	
\def\prc{\jref@jnl{Phys.~Rev.~C}}	
\def\prd{\jref@jnl{Phys.~Rev.~D}}	
\def\pre{\jref@jnl{Phys.~Rev.~E}}	
\def\prl{\jref@jnl{Phys.~Rev.~Lett.}}	
\def\pasp{\jref@jnl{PASP}}		
\def\pasj{\jref@jnl{PASJ}}		
\def\qjras{\jref@jnl{QJRAS}}		
\def\skytel{\jref@jnl{S\&T}}		
\def\solphys{\jref@jnl{Sol.~Phys.}}	
\def\sovast{\jref@jnl{Soviet~Ast.}}	
\def\ssr{\jref@jnl{Space~Sci.~Rev.}}	
\def\zap{\jref@jnl{ZAp}}			
\def\nat{\jref@jnl{Nature}}		
\def\iaucirc{\jref@jnl{IAU~Circ.}}
\def\aplett{\jref@jnl{Astrophys.~Lett.}}
\def\apspr{\jref@jnl{Astrophys.~Space~Phys.~Res.}}
\def\bain{\jref@jnl{Bull.~Astron.~Inst.~Netherlands}}
\def\fcp{\jref@jnl{Fund.~Cosmic~Phys.}}
\def\gca{\jref@jnl{Geochim.~Cosmochim.~Acta}}
\def\grl{\jref@jnl{Geophys.~Res.~Lett.}}
\def\jcp{\jref@jnl{J.~Chem.~Phys.}}	
\def\jgr{\jref@jnl{J.~Geophys.~Res.}}	
\def\jqsrt{\jref@jnl{J.~Quant.~Spec.~Radiat.~Transf.}}
\def\memsai{\jref@jnl{Mem.~Soc.~Astron.~Italiana}}
\def\nphysa{\jref@jnl{Nucl.~Phys.~A}}
\def\physrep{\jref@jnl{Phys.~Rep.}}
\def\physscr{\jref@jnl{Phys.~Scr}}
\def\planss{\jref@jnl{Planet.~Space~Sci.}}
\def\procspie{\jref@jnl{Proc.~SPIE}}

\def\ion#1#2{#1$\;${\small\rm\@Roman{#2}}\relax}

\newcount\lecurrentfam
\def\LaTeX{\lecurrentfam=\the\fam \leavevmode L\raise.42ex
\hbox{$\fam\lecurrentfam\scriptstyle\kern-.3em A$}\kern-.15em\TeX}
\def\sizrpt{
(\fontname\the\font): em=\the\fontdimen6\font, ex=\the\fontdimen5\font
\typeout{
(\fontname\the\font): em=\the\fontdimen6\font, ex=\the\fontdimen5\font
}}
\definecolor{dark}{rgb}{ 0.34, 0.09, 0.49}

\usepackage{fancyhdr,url}
\usepackage{transparent}
\usepackage{tablefootnote}
\usepackage{mathpazo}

\makeatletter
\renewcommand\section{\@startsection{section}{1}{\z@}%
  {-2.0ex \@plus -0.6ex \@minus -.2ex}{0.9ex \@plus.2ex}{\normalfont\Large\bfseries}}
\renewcommand\subsection{\@startsection{subsection}{2}{\z@}%
  {-1.6ex \@plus -0.5ex \@minus -.2ex}{0.6ex \@plus .2ex}{\normalfont\large\bfseries}}
\makeatother
\begin{document}

\noindent
\begin{raggedright}
{\Huge\bfseries ALMA2040: The High-Redshift Circumgalactic Medium\\[0.5em]
\Large Imaging the fueling and feedback of galaxies in the early universe
}\end{raggedright}\\

{\sloppy
\noindent Kirsty~M.~Butler,$^1$$^\dagger$
Justin~Spilker,$^2$$^\dagger$
Seiji~Fujimoto,$^{3,4}$$^\dagger$
Manuel~Aravena,$^5$
Cecilia~Bacchini,$^{6,7}$
Leindert~A.~Boogaard,$^{8}$
Stefano~Carniani,$^{9}$
Chiara~Circosta,$^{10}$
Pierre~Cox,$^{11}$
\mbox{Rebecca~Davies,$^{12}$}
Tanio~Diaz-Santos,$^{13}$
Daniel~Espada,$^{14}$
Andreas~Faisst,$^{15}$
Jonathan~Freundlich,$^{16}$
Mahsa~Kohandel,$^{9}$
Andrea~Lapi,$^{17,18}$
Umberto~Maio,$^{18,19}$
Dieu~D.~Nguyen,$^{20}$
Andrea~Pallottini,$^{9,21}$
Mamta~Pandey-Pommier,$^{22}$
Antonio~Pensabene,$^{23,24}$
Miguel~Pereira-Santaella,$^{25}$
Matus~Rybak,$^{26}$
Francesco~Salvestrini,$^{17,23}$
Livia~Vallini,$^{27}$
Anita~Zanella$^{27}$
\par}

{\footnotesize{
\noindent{$\dagger$ These authors contributed equally to this work}\\
\textit{Affiliations may be found after the references.}
}}


\begin{tcolorbox}[boxsep=3pt,left=0pt,right=0pt,top=1pt,bottom=1pt,colframe=boxcolor,colback=boxcolor]
%
%
%
\textbf{Abstract:} 

Galaxies do not form in isolation, but rather as part of a wider interconnected web mediated at their interface by an extended gaseous atmosphere -- the Circum-Galactic Medium (CGM). This dynamic and multiphase region is fed and shaped by gas inflows and mergers from the intergalactic medium and by feedback-driven outflows and radiation from the central galaxy. The CGM is imprinted by the past, present and future evolution of its central galaxy and environment. The cool atomic and molecular phases are of particular interest as they dominate the mass budget of outflows and provide a reservoir of direct star forming fuel for the galaxy to accrete. In the early universe where accretion rates, mergers, and feedback processes peak, the cycle of gas through the CGM is most vigorous. Sub-mm observations have detected cold gas in extended halos and gas flows around high redshift galaxies but are thus far limited to the most massive, extreme or gravitationally lensed sources, using just a handful of the strongest emission and absorption lines. A significant increase ($5-10\times$) in spectral line sensitivity compared with the current ALMA is needed to push physical scales, densities and properties to a new regime. We present three key scientific cases for an ALMA2040 telescope: (i) direct imaging of extended \cii-traced CGM halos around Milky-Way-progenitor galaxies at $z=4$--8;
(ii) resolved detection of cold, clumpy outflows on $\sim$100 pc scales tied to individual star-forming clumps; and (iii) multi-species/-transition emission and absorption diagnostics that jointly constrain the density, chemistry, and radiation fields permeating the cool CGM and its multiphase gas flows.

\end{tcolorbox}

\color{black}

\newpage{}

\vspace{-0.5cm}
\section{Scientific context and open questions}
\vspace{-0.1cm}

Astronomers once thought of galaxies as `island universes' -- separated by such vast distances that they could never interact with each other or their environments. We now know that this picture is fundamentally wrong.
Galaxies both influence and are influenced by their surroundings in vast ecosystems that govern their ability to build new stars, grow supermassive black holes, and produce (and retain) the basic elements necessary for life. Indeed, the majority of baryons within dark matter halos reside in the dynamic and multiphase Circum-Galactic Medium (CGM) \citep{Werk2014,Tumlinson2017}. The CGM represents the reservoir of gas fed by inflows from the cosmic web and outflows from the galaxy, and depleted by accretion onto the galaxy and losses to intergalactic space. As we look to early times, galaxies were more compact, had higher star formation rates (SFRs) and accretion rates from the cosmic web, and more frequently hosted powerful active galactic nuclei (AGN; \citealt{miller26,popesso23}). As a consequence, the energy and momentum injection rate available to power outflows was orders of magnitude higher in early galaxies than today, maximally impacting galaxy evolution and quenching. Detecting gas flows in high redshift systems, however, requires sensitive, high-resolution observations. 

Over the last decade, ALMA/NOEMA and JWST have revolutionised this field by detecting the extended CGM and in/outflows of cold and warm gas, respectively. JWST in particular now routinely detects outflows of warm, ionised gas even in typical high-redshift and low-mass galaxies \citep{Carniani2024}. ALMA and NOEMA, meanwhile, have detected cold neutral/molecular gas in inflows, outflows, and the extended CGM of more extreme systems \citep{Berta2021,Spilker2020,Fujimoto2019}, offering convincing evidence that the direct fuel for star formation is fundamentally connected to the feeding and feedback of the CGM. With considerable effort, these studies are now extending to the most massive of typical star-forming galaxies at $z\sim5$ \citep{ginolfi20,birkin25,HerreraCamus2025,davies26}.

While JWST forges ahead by detecting and mapping warm, diffuse, ionised gas at higher redshifts and in more typical low-mass galaxies, the cold gas that drives galaxy growth remains largely invisible, limited by the current ALMA sensitivity to more luminous galaxies. In this white paper, we argue that progress in this field requires the transformational capabilities of ALMA2040. Only a large increase in sensitivity, together with high angular resolution, will allow us to detect, image and constrain the physical properties of cold gas in the CGM and its gas flows, allowing us to answer qualitatively new questions beyond the reach of the current ALMA.

\section{ALMA2040: Key science cases for the CGM and galactic winds}

\subsection{Imaging of the Cool CGM around Normal Star-Forming Galaxies at \textit{z}~$\simeq 4-8$}\label{sec:extended}

Emission-line mapping is a powerful probe for directly imaging the cool CGM, complementing the sight-line view from absorption, where the declining C\,\textsc{iv}/C\,\textsc{ii} ratio towards the reionisation era already points to a progressively more neutral, cooler CGM \citep{DaviesRL2023}. Deep ALMA observations show that [C\,\textsc{ii}] 158\,$\mu$m emission, the brightest tracer of cool/cold gas, is systematically more extended than the rest-UV and dust continuum in star-forming galaxies at $z=4$--7 \citep{Fujimoto2020,Herrera-Camus2021,Ikeda2025,HerreraCamus2025}, reaching 10--15\,kpc scales in stacked samples \citep{Fujimoto2019,Ginolfi2020} and, in the deepest individual follow-ups, taking the form of halos, plumes, and tidal-like extensions \citep{Akins2022,Lambert2023,Solimano2024,Posses2025}. These halos show that enriched cool gas is already in place around galaxies within the first $\sim$0.8--1.5\,Gyr, but their origin remains open: star-formation-driven outflows, accreting streams, stripped satellite material, or a combination of these \citep{Pizzati2020,Pizzati2023,liu26}. Each scenario predicts a different surface-brightness profile, kinematic pattern, and metal distribution on 5--20\,kpc scales. None of these predictions can be tested on individual normal galaxies today due to limitations in surface brightness sensitivity.
Meanwhile, JWST has begun to resolve the warm ionised side of the same extended gas. NIRSpec IFU observations at $z=4$--6 map the rest-optical nebular lines on kpc scales, revealing H$\alpha$ halos beyond the stellar body \citep{Faisst2026} together with resolved diagnostics of the electron density, metallicity, and ionisation parameter across the disks \citep{RodriguezDelPino2024,Fujimoto2025,Lee2026}. Stacking of the galaxy outskirts shows that the surrounding gas at $z \simeq 5$ is already enriched to near-ISM abundances, with line ratios driven by a drop in ionisation rather than in metallicity \citep{Wang2026}. A comparable census of the cool neutral phase, which carries most of the mass of the cool gas flows, is exactly what ALMA2040 provides.

\begin{figure}[t!]
\centering
\includegraphics[width=1.0\textwidth]{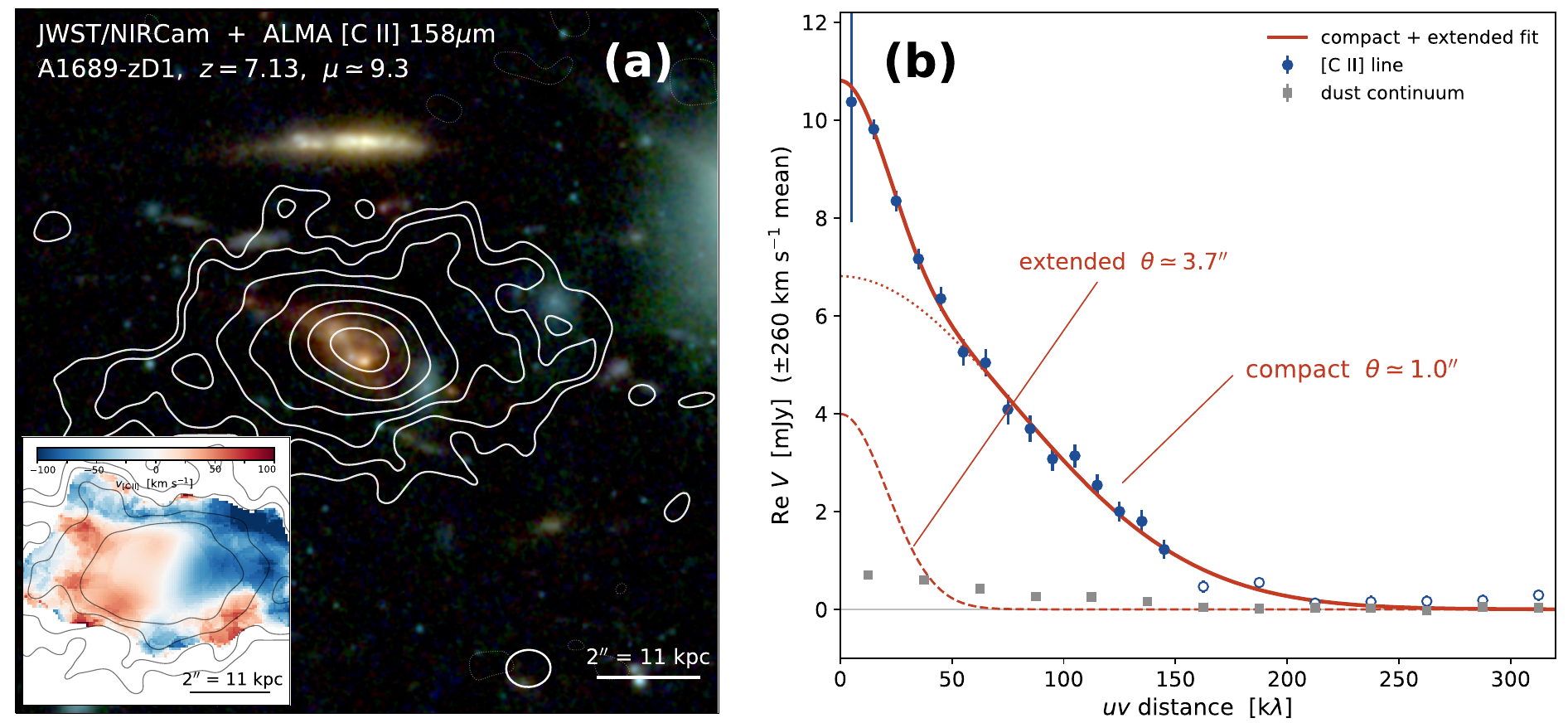}
\vspace{-0.6cm}
\caption{\it Deep ALMA view of the lensed galaxy A1689-zD1 ($z=7.13$, $\mu \simeq 9.3$; 7.8\,hr on-source in Band~6; \citealt{Watson2015,Knudsen2025}).
(a) JWST/NIRCam colour image with [C\,\textsc{ii}] 158\,$\mu$m contours (2--64$\sigma$): the cool gas extends far beyond the stellar body. The inset shows the [C\,\textsc{ii}] velocity field. (b) Real part of the [C\,\textsc{ii}] visibilities versus $uv$ distance. A two-component compact plus extended model is required to fit the \cii data while the dust (grey) stays compact, the direct interferometric signature of a cool-gas halo \citep{Akins2022}. \vspace{-0.4cm}}
\label{fig:cgm}
\end{figure}

\medskip
\noindent
{\bf A gravitationally lensed view with current ALMA.}
The lensed galaxy A1689-zD1, a typical star-forming system at $z=7.13$ with $M_\star \simeq 2 \times 10^{9}\,\msun$ \citep{Watson2015} and SFR $\simeq 37\,\msun\,{\rm yr^{-1}}$ \citep{Bakx2021,Akins2022} and magnification $\mu \simeq 9.3$ \citep{Knudsen2017}, demonstrates capabilities that will be routine with ALMA2040. Deep ALMA imaging detects extended [C\,\textsc{ii}] emission out to $r \simeq 12$\,kpc in the image plane (Fig.~\ref{fig:cgm}a), roughly four times larger than the dust continuum and the [O\,\textsc{iii}] 88\,$\mu$m line, which are each well fit by a single compact ($\simeq 1''$) component \citep{Akins2022}. The [C\,\textsc{ii}] $uv$-profile, on the other hand, requires a compact (FWHM $\sim$5\,kpc) plus an extended (FWHM$\sim$20\,kpc) component (Fig.~\ref{fig:cgm}b). The velocity field shows a coherent gradient across the emission (Fig.~\ref{fig:cgm}a, inset); whether it reflects rotation of an extended CGM gas reservoir, an interaction, or outflowing gas remains open for now \citep{Akins2022,Wong2022,Knudsen2025}. All of this, however, relies on a single fortunate alignment: the lens boosts the line sensitivity by a factor of $\sim$10, and only a handful of comparably magnified normal galaxies at $z\simeq$4--8 with successful ALMA detection exist for now. 

\medskip
\noindent
{\bf From a few lensed systems to a physical census with ALMA2040.}
Low-mass galaxies are far more numerous than the bright systems accessible today, and they are the progenitors of Milky Way-like galaxies. ALMA2040's 10$\times$ gain in line-sensitivity will match the boost from lensing provided here, such that maps of comparable quality become routine for unlensed normal-star-forming galaxies at $z\simeq4$--8, free of the distortion and differential magnification that complicate lensed observations. 
Halo mapping then moves from stacks to individual, representative galaxies, and two measurements become standard. (i) \emph{Cool-gas budgets}: surface-brightness profiles and cool-gas masses of the CGM out to $\gtrsim$10--15\,kpc, with individual maps for $\log(M_\star/\msun) \gtrsim 9.5$ galaxies in a few hours per source and the $\log(M_\star/\msun) \sim 9$ regime reached in modest sub-stacks, reaching below the knee of the stellar mass function \citep{weibel24}. This will complete the cool side of the baryon budget as a function of mass and redshift and directly test outflow versus accretion origins \citep{Pizzati2020,Pizzati2023}. (ii) \emph{CGM kinematics}: resolved velocity and line-width maps that help to distinguish outflowing shells, extended rotating reservoirs, and accreting gas. Enriched, recycled inflows appear in emission, while pristine streams are constrained jointly with the absorption sight lines discussed in Section~\ref{sec:smallscale}; together with the clump-scale outflows of Section~\ref{sec:clumps} these fix the mass and momentum loading of the baryon cycle. 
Separating rotation, outflow, and inflow may take more than a first-moment map, where forward-modelling the beam-convolved cube isolates the rotating component \citep[e.g.,][]{DiTeodoro2015,Davis2013,RomanOliveira2023}. 
Typical S/N in current high-$z$ studies provide a biased recovery of parameters, such as rotation velocity and velocity dispersion \citep{Yttergren2025}, however, per-object decomposition on halo scales, possible with ALMA2040's 5-10$\times$ sensitivity gain, will solve this. Emission alone leaves the sign of radial motion open, but is settled using complementary absorption sight lines. 
Matched maps of [C\,\textsc{ii}] and fainter FIR lines then turn these images into densities, masses, and a cooling budget for the enriched gas (Section~\ref{sec:smallscale}), in direct synergy with the JWST/NIRSpec view of the warm ionised phase \citep{Faisst2026,Wang2026}. 

\subsection{Resolving cold outflows in the early universe} \label{sec:clumps}

While the gas in outflows spans many orders of magnitude in both temperature and density, it is the cold, dense, and neutral atomic / molecular phase that is thought to dominate the total mass budget \citep[e.g.][]{Fluetsch2019}. High-resolution simulations predict that the properties of outflows depend strongly on the small-scale local properties of the regions that launch them, such as the local SFR surface density \sigsfr \citep[e.g.][]{kim20,rathjen23}. The cold gas phase is the most challenging for simulations to tackle. In the classical cloud-crushing picture, directly accelerating cold gas via ram pressure from the hot, fast winds driven by supernovae or AGN accretion disks tends to shred and destroy the cold gas long before it reaches the $\approx$1000\,\kms velocities we observe. This has led theorists to speculate that, for example, magnetic fields or thermal conduction can protect the cold gas long enough to accelerate \citep[e.g.][]{armillotta17,anglesalcazar21}. Alternatively, perhaps the molecular gas in outflows was not molecular when it was accelerated but instead cooled and condensed from hotter gas in situ in the CGM \citep[e.g.][]{richings18,costa20}, implying a much less direct connection to the regulation of star formation in galaxies. These scenarios result in large, qualitative differences in the small-scale structure of the cold outflowing gas, ranging from a `mist' of cold clumps on $\sim$10\,pc scales condensed from the hot phase to much larger 100--500\,pc-scale dense knots that survive direct acceleration.

Cold outflows are much more difficult to detect than warmer gas phases observed in the rest-optical, and consequently the connection between the cold outflowing gas and the local conditions of the regions that launch winds is poorly understood. A large fraction of high-redshift galaxies have irregular, clumpy morphologies, with most of the galaxy's star formation taking place in a few kiloparsec-scale giant clumps \citep{elmegreen09,wuyts12,huertascompany24}. Clumps form in regions where the galactic disk is gravitationally unstable to collapse \citep{genzel11}, and appear to be key sites where winds are launched by the local high \sigsfr \citep{murray11,kalita24}. Connecting outflows to the star-forming clumps associated with them remains observationally challenging, requiring both high line sensitivity and high physical resolution.

\sloppy The recent CRISTAL ALMA Large Program \citep{HerreraCamus2025} illustrates the challenges of identifying cold outflows and connecting them to the kpc-scale regions that drive them with the current capabilities of ALMA. CRISTAL observed 25 ALMA pointings containing 39 galaxies at $z\sim4.5-5.5$ with a total telescope time of $\approx$200\,hours (including archival datasets). The sample focused on galaxies at the massive end of the star-forming main sequence, $\log(\Mstar/\Msun) \gtrsim 9.5$ (corresponding to SFR $\gtrsim$ 20\,\Msun/yr), targeting the redshifted \cii 158\,\um line at typical $\approx$2\,kpc resolution. By stacking the integrated spectra, \citet{birkin25} detected high-velocity \cii line wings that plausibly represent outflowing gas, but the line wings were $\sim3\times$ weaker than in a similar analysis using low-resolution ALMA data \citep{ginolfi20}. This is plausibly because the earlier data were unable to excise satellite galaxies or mergers, where tidal interactions can also result in high-velocity gas \citep{birkin25}. High resolution data allows these interactions to be resolved, and allows lower-speed outflows to be recovered because the local systemic velocity can be estimated in each pixel, instead of from only a total spectrum. 

\begin{SCfigure}[][h]
\centering
\includegraphics[width=0.7\textwidth]{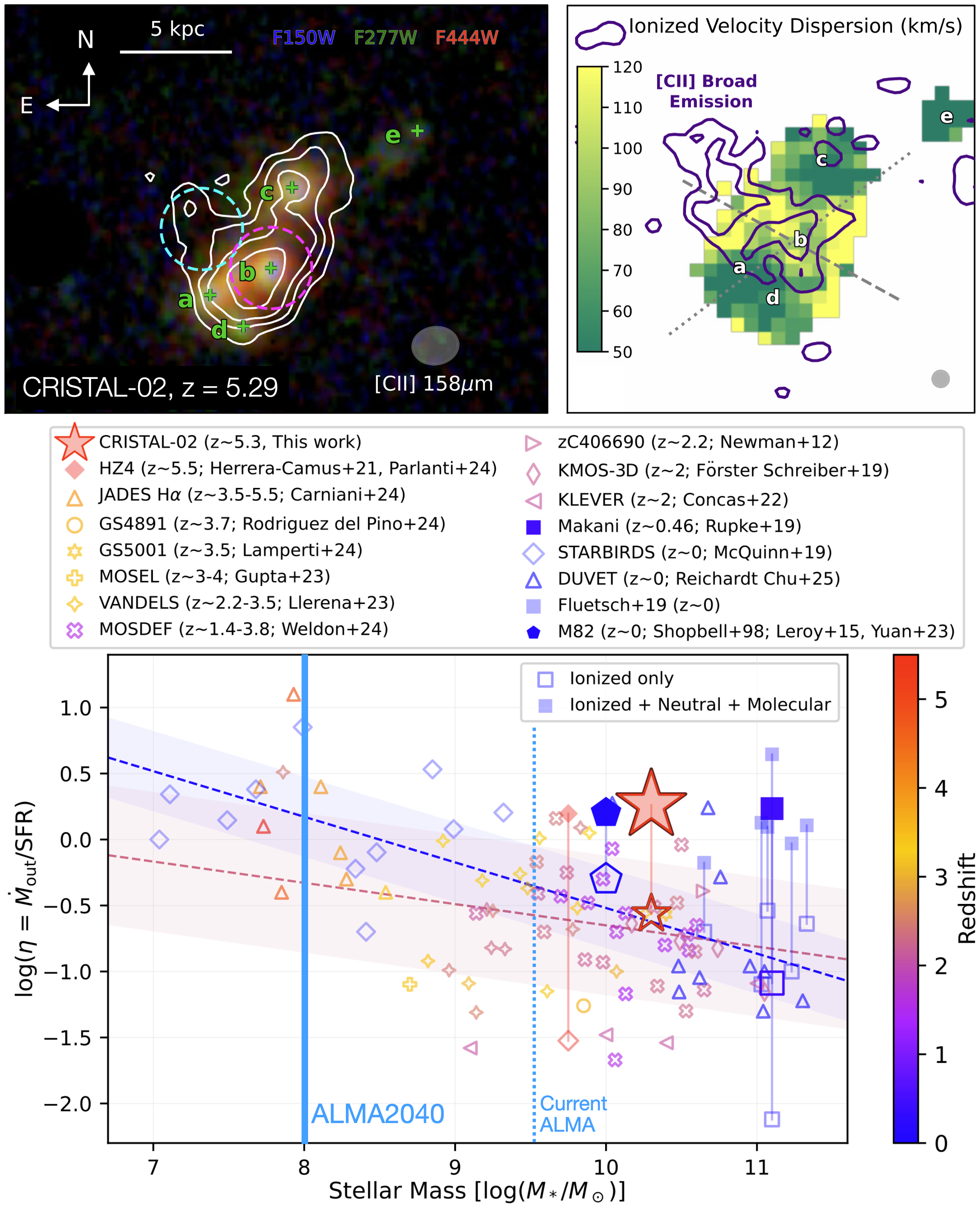}
\caption{
High line sensitivity and resolution are required to detect and resolve cold outflows at $z\sim5$, illustrated here in the galaxy CRISTAL-02 from the CRISTAL Large Program \citep{davies26,HerreraCamus2025}. JWST imaging and IFU spectroscopy resolve the galaxy into 5 star-forming clumps, `a--e' (upper left). Only Clump `b' shows signs of an outflow in cold (\cii 158\,\um) or warm gas (H$\alpha$; upper right), and the cold gas phase dominates the total outflow mass (lower panel), similar to archetypal local-universe winds. ALMA2040 will reach $\approx$30$\times$ lower-mass galaxies that are far more numerous than this case study. Reproduced from \citet{davies26}.
}\label{fig:davies26}\vspace{0.15in}
\end{SCfigure}

Figure~\ref{fig:davies26} shows the benefit of high resolution directly in a case study of the galaxy CRISTAL-02, one of the highest-SFR targets in the CRISTAL sample \citep{davies26}. JWST imaging and IFU spectroscopy resolves the galaxy into at least 5 star-forming clumps, but only Clump `b' appears to drive an outflow, visible as an extended plume of \cii emission and an increase in the local velocity dispersion of both \cii and H$\alpha$. Joint analysis of the JWST and ALMA data show that the cold gas traced by \cii dominates the warm ionised gas by $\approx$0.8\,dex -- JWST spectroscopy alone provides only a highly incomplete view of the clump-scale feedback (Fig.~\ref{fig:davies26} lower panel). This example demonstrates that resolved studies of cold outflowing gas are at the outer edge of the current ALMA capabilities. The CRISTAL sample consists of very massive, highly star-forming galaxies, and required the substantial investment of a Large Program to detect cold outflows in $\approx$2\,kpc resolution data. Detecting outflows in less massive systems and resolving the outflow structure on the relevant $\sim$100\,pc scales predicted by simulations are well beyond the current ALMA capabilities due to low sensitivity, both generally and on $<$0.05'' angular scales.

 Clearly a facility like ALMA2040 would make such large investments nearly trivial, allowing far larger samples and a statistical handle on the cold outflow properties of  high-redshift galaxies. The 10$\times$ increase in line sensitivity from ALMA2040 is critical to make progress in this area, enabling two qualitatively new advances over the existing ALMA: (i) \textit{A census of cold outflows in faint normal galaxies:} Existing programs like CRISTAL are limited to massive and highly star-forming galaxies by the line sensitivity of ALMA, but outflow properties are expected to scale strongly with host galaxy properties. In particular, most simulations predict that the outflow mass loading increases in lower-mass galaxies as $\approx \Mstar^{-(1/3\,\rm{to}\,2/3)}$ \citep[e.g.][]{nelson19,wright24}. Thus (for example) a 10$\times$ gain in line sensitivity from ALMA2040 will bring the cold outflow phase into view for galaxies $\sim$30--50$\times$ less massive than what ALMA can presently achieve (see Fig.~\ref{fig:davies26} lower panel). (ii) \textit{Small-scale structure of cold outflows:} Existing ALMA programs can marginally resolve cold gas outflows to localise their launch sites in the most massive star-forming galaxies at $z\sim5$ \citep[e.g.][]{davies26}, but their small-scale structure remains out of reach. From direct observation in the local universe and optical depth arguments at earlier times, cold outflows have substructures on $\approx$100\,pc scales \citep{Spilker2020,bolatto24}. Directly resolving this substructure is out of reach of the existing ALMA capabilities, even with the benefit of gravitational lensing magnification. 

\subsection{Physical and chemical Properties of the Cool CGM and its Gas Flows} \label{sec:smallscale}

The physical and chemical properties of the CGM are sensitive to both the processes occurring within the central galaxy and its wider environment. At the same time, the properties of the CGM impact the timescales over which gas can be accreted or reaccreted by the galaxy, thereby playing a key role in the regulation of galaxy growth and quenching. The atomic and molecular components are of particular interest as they dominate the mass flux of in- and outflowing gas to the galaxy and constitute the direct fuel for star formation. Thus, the distribution of gas phases, densities, temperatures, and radiation fields permeating the cool CGM and its gas flows encode important information about the past, present, and future evolution of its central galaxy. Determining these physical properties, however, requires multi-line, multi-species observations that, in turn, require extremely sensitive facilities to be detected at high redshift.

Evidence of neutral $T\sim10^4$\ K cool gas in the CGM of high-redshift galaxies came at the turn of the century thanks to bright Ly$\alpha$ emission \citep{Hayashino2004}. However, extracting the properties of the gas traced by Ly$\alpha$ is strongly model dependent and difficult to interpret due to its resonant nature. Instead, redshifted FIR spectral lines detected by powerful sub-mm telescopes such as NOEMA and ALMA have since provided crucial information of the extended cool gas halos and embedded gas flows around high-redshift galaxies \citep{Fujimoto2019,Spilker2020}. Despite this success, the library of molecular and atomic species known to reliably trace these faint structures is small and generally limited to only the strongest emission (\cii 158\,\um) and absorption lines (OH 119\,\um, $\rm OH^+\ 1_1-0_1$, $\rm CH^+\ 1-0$). Consequently, the majority of high-redshift CGM studies targeting cool gas employ just a single, redshift-dependent tracer, providing only a partial view of the cool gas reservoir and its properties. A significant leap in spectral line sensitivity compared to the current ALMA is needed to bring fainter FIR lines above our detection limit, allowing us to determine the physical conditions of high redshift CGMs.

Every spectral line or chemical species probes a unique distribution of gas properties and by themselves reveal either a restricted and/or overly general view of the total gas reservoir. For example, the bright \cii 158\,\um emission line, which has been instrumental in revealing extended CGM reservoirs and outflowing gas around high-redshift galaxies (Sec. \ref{sec:extended} and \ref{sec:clumps}), broadly emits across molecular, atomic, and ionised gas phases. To determine the relative contribution of these phases, additional lines and species are needed. In emission, promising results have emerged using low-J rotational lines of CO, with ALMA CO\ 4-3 observations revealing extended molecular gas reservoirs out to 100 kpc in a handful of the most massive star-forming and ultraluminous active high redshift galaxies \citep{Ginolfi2017,Li2023}. In CO-poor molecular gas (e.g., due to metallicity, cosmic ray ionisation ect.), \ci observations can fill in these gaps but is currently detected in very few high-redshift CGMs \citep[e.g.][]{Ghodsi2024}. Theoretical work has highlighted a potentially significant contribution from the CGM to \oiii\ emission in high-redshift sources, which traces warm ionised gas directly surrounding the galaxy and is distinct from the reservoirs traced by CO, \ci and \cii \citep{Schimek2024}.

Absorption lines of light hydride molecules (one heavy element atom attached to one or more hydrogen atoms), observed against a galaxy's own dust continuum, have provided a complimentary view of the neutral CGM at high redshift. Since the chemical pathways for producing light hydrides in diffuse gas are simple, the analysis of their observed abundances is relatively straightforward \citep{Gerin2016}, making them powerful probes of the surrounding physical properties. $\rm CH^+$ for example traces the dissipation of mechanical energy in diffuse, partially molecular gas, allowing their line properties to infer the sizes of turbulent high-redshift gas halos \citep{Falgarone2017}. Cosmic Rays (CR) are responsible for kicking off the production routes of many light hydride species, and are of particular interest in CGM studies as CR pressure acts to increase cloud buoyancy and regulate thermal instabilities, allowing cold gas to survive for longer. This pushes cold gas to larger radii \citep{Hopkins2020,Ji2020} and increases the free fall time for gas accreting or re-accreting through the CGM, significantly impacting the fuel supply of the central galaxy \citep{Su2020}. $\rm OH^+$ is one such molecule that requires CR ionisation of neutral hydrogen to form in the diffuse neutral medium. It has been observed towards many high-redshift star-forming \citep{Riechers2021} and active \citep{Butler2023} galaxies, but must be analysed in combination with other neutral gas and CR-sensitive molecules, such as $\rm H_2O^+$, to estimate the CR ionisation rate and remove the degeneracy with molecular gas fraction. Absorption depths of the strongest $\rm H_2O^+$ transitions, however, are generally ~25\% that of $\rm OH^+$, and so such studies only exist for a few strongly lensed high-redshift sources \citep{Indriolo2018}. A $10\times$ increase in line sensitivity will allow the same science in samples of unlensed sources and across a broader range in SFR of lensed sources. At the same time, other more constraining species such as $\rm ArH^+$ \citep{Jacob2022} will become feasible for significant samples of lensed sources at high-redshift.

As with extended CGM reservoirs, a comprehensive view of the multiphase outflowing gas is needed at high-redshift. The species and line combinations mentioned above are all also relevant for constraining the total mass and multiphase physical properties of the outflowing gas. The gas density distribution provides insights into outflow driving mechanisms \citep[e.g.][]{Fluetsch2019} and their survival through the turbulent hot CGM towards possible re-accretion. Currently missing from high-redshift outflow observations are species that trace the densest molecular and most diffuse atomic phases. Redshifted $\rm H_2O\ 4_{2,3}-4_{1,4}$ and $3_{3,0}-3_{2,1}$ absorption, tracing warm very dense gas has been observed in one high-redshift gravitationally lensed hyper-starburst galaxy \citep{Jones2019} thanks to its $5\times$ magnification. At even higher critical densities, molecules such as HCN, HNC etc. provide the ideal tracers for the densest molecular gas, but are typically 10$\times$ or more fainter than bulk gas tracers (CO, \cii). Dense gas observations in $z\geq1$ dusty star-forming galaxies and QSOs are therefore limited and have provided evidence of outflowing gas in just one high-redshift, gravitationally lensed dusty star-forming galaxy from a sample of 11, revealed by a P-Cygni profile in the HNC\ 4-3 transition \citep{Rybak2026}. A $10\times$ increase in line sensitivity will allow dense gas detections in unlensed starburst galaxies across cosmic noon, thereby providing greatly expanded samples from which dense gas outflows can be sought. Since dense gas typically traces the most compact and clumpy structures in galaxy outflows (down to pc scales; \cite{Otter2026}), the high spatial resolution provided by an ALMA-2040 interferometer will be essential for avoiding beam dilution likely present in current unresolved observations (Sec. \ref{sec:clumps}). At the other end of the spectrum, $\rm ArH^+$ absorption offers a promising tracer of the purest diffuse atomic gas phase \citep{Schilke2014}, but is currently only observable at reasonable integration times in the brightest gravitationally lensed dusty high-redshift sources. Across the board, there is a need for significant increases in spectral line sensitivity to transition high-redshift outflow studies from simple bulk estimates to comprehensive and detailed multiphase science.

\section{ALMA2040: Technical Requirements}

Transformative advances in our understanding of the high-redshift cold CGM and gas flows will require a next-generation interferometer with 10$\times$ higher line sensitivity compared to existing ALMA across the prime sub/millimetre atmospheric windows. This science case simultaneously requires excellent surface-brightness sensitivity to diffuse emission on $\gtrsim$10 kpc scales and $\sim$100 pc resolution for compact outflow structures. This will likely require a large increase in collecting area over ALMA, coupled with advances in receiver sensitivity. 

Increased line sensitivity is the primary requirement to detect the cold CGM and gas flows in the typical low-mass galaxies that evolve to become Milky Way-like by $z=0$. Increasing sensitivity is more costly at higher frequencies (e.g. requiring more accurate antenna shapes). The science cases presented here require a $\sim$10$\times$ gain up into the submillimetre windows, at least ALMA Bands 7 and 8; relaxing this requirement limits the redshift ranges over which the key spectral lines can be observed homogeneously.

Detailed inventories of outflows and the CGM benefit from an ability to detect multiple spectral lines simultaneously. The rest-frame far-IR atomic and molecular lines that are key diagnostics of the physical properties and chemistry of the cold gas are widely separated in frequency; a full inventory could thus be achieved either with extremely wide continuous spectral coverage or narrower, flexibly tuned spectral windows separated by tens of GHz.

This science case does not necessarily require longer baselines than can be achieved in the existing ALMA concession; the current limitation is primarily surface brightness sensitivity. Short baseline coverage will be necessary to detect extended CGM gas. For 16\,km baseline observations, ALMA2040 will reach $\approx$100\,pc physical scales when observing rest-frame far-IR lines (\cii 158\,\um, OH 119\,\um, etc.) across $z\sim3-8$. At larger scales, combined observations with the ALMA2040 Total Power single-dish array ($5\times24$\,m dishes) will be essential. Detailed trade-off studies of baseline distribution, and thus surface brightness sensitivity as a function of angular resolution, will be needed to understand the detectability of both diffuse and small-scale clumpy outflow and CGM structures.

\section{Synergies with current and future facilities}

A comprehensive understanding of the CGM and galactic winds is an inherently multiwavelength proposition, since the multiphase gas spans many orders of magnitude in both temperature and density. ALMA2040 will revolutionise our view of cold gas properties and structures in the CGM, tracing the gas most directly related to star formation. 

The Square Kilometre Array (SKA) and Next Generation Very Large Array (ngVLA) will provide complementary views of the cold phase of the CGM. The ngVLA will detect outflows in the ground-state transitions of many molecular species in typical main-sequence galaxies, including molecular transitions that trace gas at densities far higher than the ions and molecules accessible to ALMA2040 (CO, HCN, HNC, CS, etc.). The SKA, meanwhile, can directly observe the neutral atomic hydrogen 21\,cm transition in absorption in galaxies that host even modest levels of radio AGN activity. Combining SKA HI measurements with ALMA2040 observations of \cii, CO and other molecular tracers could therefore provide complementary constraints on the atomic and molecular phases of the cold gas cycle. Where sufficiently bright background radio continuum sources are available, HI absorption would additionally provide a kinematic diagnostic that complements ALMA2040 emission measurements and can help constrain the nature of inflowing, outflowing and dynamically disturbed gas \citep{Mahony2026,PandeyPommier2026}. Similarly, NASA's recently selected PRobe far-Infrared Mission for Astrophysics (PRIMA) will provide crucial access to hydrocarbon dust features, allowing studies of the kinematics and chemical composition of dust present in high redshift molecular outflows \citep{Moullet2025}.

At warmer $\sim$10$^4$-10$^6$\,K temperatures, the completion of the next generation of ground-based Extremely Large Telescopes (ELTs) in the early/mid-2030s and the 2040s-era launch of the Habitable Worlds Observatory (HWO) will advance this field in two key ways \citep{burchett25}. Firstly, the extremely high angular resolution of these facilities will result in maps of galaxies that detect and resolve any and all star-forming clumps, if present, on scales down to $\approx$200\,pc. It is critical that the long-wavelength community be able to reach similar resolution with ALMA2040 at high surface brightness sensitivity. Second, integral-field spectrographs on these facilities will image the warm CGM gas in emission by mapping Ly$\alpha$ and UV metal lines that trace a wide range of temperature and ionisation states (e.g. Mg\,II, Si\,II, Si\,III, C\,III, C\,IV, O\,IV, O\,VI).

The hottest gas phases ($\gtrsim$10$^6$--10$^7$\,K) are most directly detectable in the X-ray. This extremely hot plasma fills most of the volume of the CGM and carries away the largest portion of the thermal and kinetic energy of outflows, but is also very tenuous. At present, the 2040s-era landscape of X-ray observatories is unclear, but imaging the hot CGM has been a key science case for X-ray missions proposed to space agencies around the world (e.g. Athena, Lynx, AXIS; \citealt{barret18,reynolds23}). Combining ALMA2040 observations of the cold gas, which likely dominates the mass of galactic winds, with X-ray detections of the hot gas phase, dominating the energy budget of winds, will piece together a comprehensive view of these challenging multiphase phenomena.

\section*{Acknowledgements}
This work was developed as part of the broader ALMA2040 community effort. We acknowledge the Lorentz Center and NOVA for supporting the community workshops and coordination activities that helped initiate and shape the ALMA2040 White Paper process. The authors would like to thank the internal and external reviewers for their valuable input.\\

\bibliographystyle{apj_mod}
{\bibliography{bibliography}}

\section*{Affiliations}

{
\noindent $^{1}$ Department of Physics and Astronomy, Chalmers University of Technology, SE-412 96 Gothenburg, Sweden \\ 
$^2$ Department of Physics and Astronomy and George P. and Cynthia Woods Mitchell Institute for Fundamental Physics and Astronomy, Texas A\&M University, 4242 TAMU, College Station, TX 77843-4242, US \\ 
\noindent $^{3}$ David A. Dunlap Department of Astronomy and Astrophysics, University of Toronto, 50 St. George Street, Toronto, Ontario, M5S 3H4, Canada \\ 
\noindent $^{4}$ Dunlap Institute for Astronomy and Astrophysics, 50 St. George Street, Toronto, Ontario, M5S 3H4, Canada \\ 
\noindent $^{5}$ Instituto de Estudios Astrofisicos, Universidad Diego Portales, Av. Ejercito 441, Santiago, Chile \\ 
\noindent $^{6}$ DARK, Niels Bohr Institute, University of Copenhagen, Jagtvej 155, 2200 Copenhagen, Denmark \\ 
\noindent $^{7}$ European Southern Observatory (ESO), Karl-Schwarzschild Str. 2, 85748 Garching, Germany \\ 
\noindent $^{8}$Leiden Observatory, Leiden University, PO Box 9513, NL-2300 RA Leiden, The Netherlands \\ 
\noindent $^{9}$ Scuola Normale Superiore, Piazza dei Cavalieri 7, 56126 Pisa, Italy \\ 
\noindent $^{10}$ Institut de Radioastronomie Millimétrique (IRAM), 300 rue de la Piscine, 38400 Saint-Martin-d’Hères, France \\ 
\noindent $^{11}$ Sorbonne Université, UPMC Paris 6 and CNRS, UMR 7095, Institut d’Astrophysique de Paris, 98b bd. Arago, 75014 Paris, France \\ 
\noindent $^{12}$ Centre for Astrophysics and Supercomputing, Swinburne University of Technology, John Street, Hawthorn, VIC 3122, Australia\\ 
\noindent $^{13}$ Foundation for Research and Technology--Hellas (FORTH), Heraklion, GR-70013, Greece \\ 
\noindent $^{14}$ Dpto. de Física Teórica y del Cosmos, Edificio Mecenas, Campus de Fuentenueva, Universidad de Granada, E-18071, Granada, Spain \\ 
\noindent $^{15}$ IPAC, California Institute of Technology, 1200 E. California Blvd. Pasadena, CA 91125, USA, \\ 
\noindent $^{16}$ Université de Strasbourg, CNRS, Observatoire astronomique de Strasbourg, UMR 7550, F-67000 Strasbourg, France \\ 
\noindent $^{17}$ SISSA, Via Bonomea 265, 34136 Trieste, Italy \\ 
\noindent $^{18}$ IFPU -- Institute for Fundamental Physics of the Universe, Via Beirut 2, I-34014 Trieste, Italy \\ 
\noindent $^{19}$ INAF -- Osservatorio Astronomico di Trieste, Via G. Tiepolo 11, 34143 Trieste, Italy \\ %
\noindent $^{20}$ Department of Astronomy, University of Michigan, 1085 South University Avenue, Ann Arbor, MI 48109, USA \\ 
\noindent $^{21}$ Dipartimento di Fisica ``Enrico Fermi'', Università di Pisa, Pisa, Italy \\ 
\noindent $^{22}$ Pole Scientific, University Catholic of Lyon- University of Lyon, 10 place des Archives 69002, Lyon, France \\ 
\noindent $^{23}$ Cosmic Dawn Center (DAWN), Denmark \\ 
\noindent $^{24}$ DTU Space, Technical University of Denmark, Elektrovej 327, DK2800 Kgs. Lyngby, Denmark \\ 
\noindent $^{25}$ Instituto de F\'isica Fundamental, CSIC, Calle Serrano 123, 28006 Madrid, Spain \\ 
\noindent $^{26}$ Institute for Astronomy, Royal Observatory, Blackford Hill, Edinburgh EH9 3HJ, United Kingdom \\
\noindent $^{27}$ INAF -- Osservatorio di Astrofisica e Scienza dello Spazio, Via P. Gobetti 93/3, I-40129, Bologna, Italy \\ 
}

\end{document}